\documentclass[epj]{webofc}
\usepackage[varg]{txfonts}   
\usepackage{graphicx,color} 
\usepackage{bm}       
\usepackage{amsmath}  
\usepackage{amssymb}
\usepackage{mathtools}
\usepackage{braket}

\woctitle{21st International Conference on Few-Body Problems in Physics}
\begin{document}
\title{Theory of electromagnetic reactions in light nuclei}

%
%
\author{Tianrui Xu\inst{1,2}
        \and
        Mirko Miorelli\inst{1,2} \and
        Sonia Bacca\inst{1,3}\fnsep\thanks{\email{bacca@triumf.ca}} \and
        Gaute Hagen\inst{4,5}
}

\institute{TRIUMF, 4004 Wesbrook Mall,
Vancouver, British Columbia, Canada V6T 2A3
\and
           Department of Physics and Astronomy, University of British Columbia,
Vancouver, British Columbia, Canada V6T 1Z4
\and
           Department of Physics and Astronomy, University of Manitoba,
Winnipeg, Manitoba, Canada R3T 2N2
\and
        Physics Division, Oak Ridge National Laboratory,
Oak Ridge, Tennessee 37831, USA
\and
        Department of Physics and Astronomy, University of Tennessee,
Knoxville, Tennessee 37996, USA
          }

\abstract{We briefly review the theory for  electromagnetic reactions in light nuclei based on the coupled-cluster formulation of the Lorentz integral transform method. Results on photodisintegration reactions of $^{22}$O and $^{40}$Ca are reported on and preliminary calculations on the Coulomb sum rule for $^4$He are discussed.}
\maketitle
\section{Introduction}
\label{intro}
The investigation of nuclear reactions from first principles is fundamental
in bridging  nuclear physics  with the underlying quantum chromo-dynamics regime. 
Nuclear reactions induced by electromagnetic probes are very useful as 
the electromagnetic current is well known and a clean comparison with experimental 
data can be performed~\cite{review}.
Nowadays  this valuable information is not only accessible for the lightest nuclei, 
but novel theoretical approaches are being developed to tackle nuclei with a larger number of nucleons.
Below, we briefly review the Lorentz integral transform (LIT) approach in its coupled-cluster (CC) theory formulation~\cite{PRL2013}. Thereafter, we discuss some recent results regarding photoabsorption and electron-scattering reactions.

\section{Theoretical Formulation}
\label{theory}
The key ingredient to study reactions induced by electromagnetic external probes, like photons or electrons, is the nuclear response function 
\begin{equation} 
R(\omega,q)=\sum_n \left|\bra{n}{\mathcal O}(q)\ket{0}\right|^2\delta\left(E_n-E_0-\omega \right),
\label{eq:rs}
\end{equation}
where $\mathcal O(q)$ is the excitation operator, which will depend specifically on the external probe and on the momentum-transfer $q$. The nuclear response function is a dynamical observables which requires knowledge on the whole spectrum of the nucleus, being $|0\rangle $ and $|n \rangle$ the ground- and excited-state, respectively.
The exact solutions of the many-body problem for the excited-states, typically in the continuum, is limited in mass number and in energy $\omega$. In particular, it is difficult to calculate when many channels are open. However, one can use the LIT approach~\cite{review_LIT} to reduce the continuum problem to the solution of a bound-state equation. 
In particular, recently this method was reformulated using coupled-cluster theory~\cite{shavittbartlett2009}, obtaining a new approach to extend {\it ab-initio} studies of electromagnetic observables to heavier and neutron-rich nuclei~\cite{PRL2013,PRC2014}.

In coupled-cluster theory one introduces the similarity transformed
Hamiltonian 
\begin{equation}\label{hbar}
 \overline{H} = \exp(-T) \hat H_{N} \exp(T),
\end{equation}
where $H_{N}$ is normal-ordered with respect to a chosen Slater determinant $ | \Phi_0 \rangle$.
Correlations are introduced through the cluster operator $T$
which is expanded into particle-hole ($ph$) excitation
operators, {\it i.e.}, $T=T_1+T_2+\ldots $, with the $1p$-$1h$ excitation
operator $T_1$, the $2p$-$2h$ excitation operator $T_2$, and so on.
In the coupled-cluster formalism \cite{PRL2013} the response function in Eq.~(\ref{eq:rs}) becomes
\begin{equation}\begin{split}\label{respfunc}
R(\omega, q) = \sum_n &\langle 0_L|\bar{\Theta}^\dag (q) |n_R\rangle\langle n_L|\bar{\Theta} (q)|0_R\rangle\delta(E_n - E_0 - \omega),
\end{split}\end{equation}
where $\bar{\Theta}^{(\dag)}(q)=e^{-\hat{T}}{\mathcal O}^{(\dag)} (q) e^{\hat{T }}$  is the similarity transformed (adjoint) excitation operator
 and $\langle 0_L|$, $|0_R\rangle$ ($\langle n_L|$, $|n_R\rangle$) are the left and right reference ground-states (excited-states), respectively (see also Ref.~\cite{mirko_fb21}). 
The utilization of the LIT method requires to find the solution $\tilde{\Psi}_R$  of the following bound-state equation~\cite{PRL2013,PRC2014}
\begin{equation}\label{LITCC_eq}
(\overline{H} - \Delta E_0)|\tilde{\Psi}_R\rangle = \overline{\Theta}(q)|0_R\rangle,
\end{equation}
where $\Delta E_0$ is the ground-state correlation energy and $|0_R\rangle=| \Phi_0 \rangle$.
By assuming that
$|\widetilde{\Psi}_R\rangle ={\cal R} | \Phi_0 \rangle $ and expanding  $\mathcal R$ in $ph$ excitation operators, similarly to what done with $T$, one can see that Eq.~(\ref{LITCC_eq}) is equivalent to an equation-of-motion~\cite{shavittbartlett2009} with a source in the right-hand-side. In Ref.~\cite{PRL2013,PRC2014} we have implemented this method with a truncation of both the operators $T$ and ${\mathcal R}$ up to the $2p$-$2h$ excitation level, namely up to single and double (SD) excitations. Such a scheme is being named here the LIT-CCSD method.

By solving Eq.~(\ref{LITCC_eq}) one finds the LIT of the response function, integrated from threshold $\omega_{th}$ to infinity
\begin{equation} \label{lorenzo}
  { L}(\omega_0,\Gamma )=\frac{\Gamma}{\pi}\int_{\omega_{th}}^{\infty} d\omega \frac{R(\omega,  q)}{(\omega -\omega_0)
               ^2+\Gamma^2} = \langle \tilde{\Psi}_L |\tilde{\Psi}_R \rangle \,,
\end{equation} 
where $\omega_0$ and $\Gamma$ are auxiliary parameters and $\langle \tilde{\Psi}_L|$ is the solution of the equivalent to Eq.~(\ref{LITCC_eq}) for the left-hand-side.
To obtain the response function $R(\omega, q)$ we then invert Eq.~(\ref{lorenzo})
  using the method
outlined in Refs.~\cite{efros1999,andreasi2005}, which looks for the
regularized solution of the integral transform equation. 

Below we highlight some of our recent calculations obtained for photoabsorption and electron scattering reactions using as only ingredient an Hamiltonian $H$ that contains, besides kinetic energy, a two-nucleon force derived in chiral effective field theory at next-to-next-to-next-to-next-to-leading (N$^3$LO) order~\cite{Entem03}.

\section{Photodisintegration reactions}

Photodisintegration reactions have been widely studied in the '70s with experiments on a variety of stable nuclei. The main observed feature of the measured cross sections
was a very pronounced peak, referred to as the giant dipole resonance,  located at excitation energies of about 10--30~MeV. First theoretical interpretations
 were given in terms of collective models~\cite{GoT48, steinwedel1950} and until very recently, most microscopic calculations were obtained in terms of phenomenological approaches and interactions, see, {\it e.g.}, Ref.~\cite{Erler2011}.  
The LIT-CCSD approach introduced above, offers the opportunity to investigate photodisintegration reactions from
an {\it ab-initio} point of view using nucleon-nucleon potentials, which reproduce two-nucleon scattering data~\cite{Entem03}. Moreover, it allows one to study equivalent processes in neutron-rich nuclei, which have been more recently investigated with Coulomb excitations experiments at the rare isotope facilities, see {\it e.g.} Ref.~\cite{leistenschneider2001}.

In the long-wavelength limit, the photodisintegration cross section can be written as
\begin{equation}
\label{cs_siegert}
\sigma_{\gamma}(\omega)=4\pi^2\alpha~\omega~R^{D}(\omega)\,, 
\end{equation}
where $\alpha$ is the electromagnetic coupling constant and $R^{D}(\omega)$ is the dipole response function, namely Eq.~(\ref{eq:rs}) with the  the dipole operator as  excitation operator 
\begin{equation}
\label{dipole}
D_z=\sum_k^A \left({z}_k -{Z}_{cm} \right) \left (\frac{1+\tau^3_k}{2} \right)\,.
\end{equation}
Here, ${z}_k$ is the $z$-coordinate of the $k$-th nucleon in the lab-frame, while
${Z}_{cm}$ denotes the $z-$component of the center of mass of the nucleus and $\tau^3_k$ is the third component of
the isospin of the $k$-th nucleon.

\begin{figure}
\centering
\begin{minipage}{.46\linewidth}
  \includegraphics[width=\linewidth]{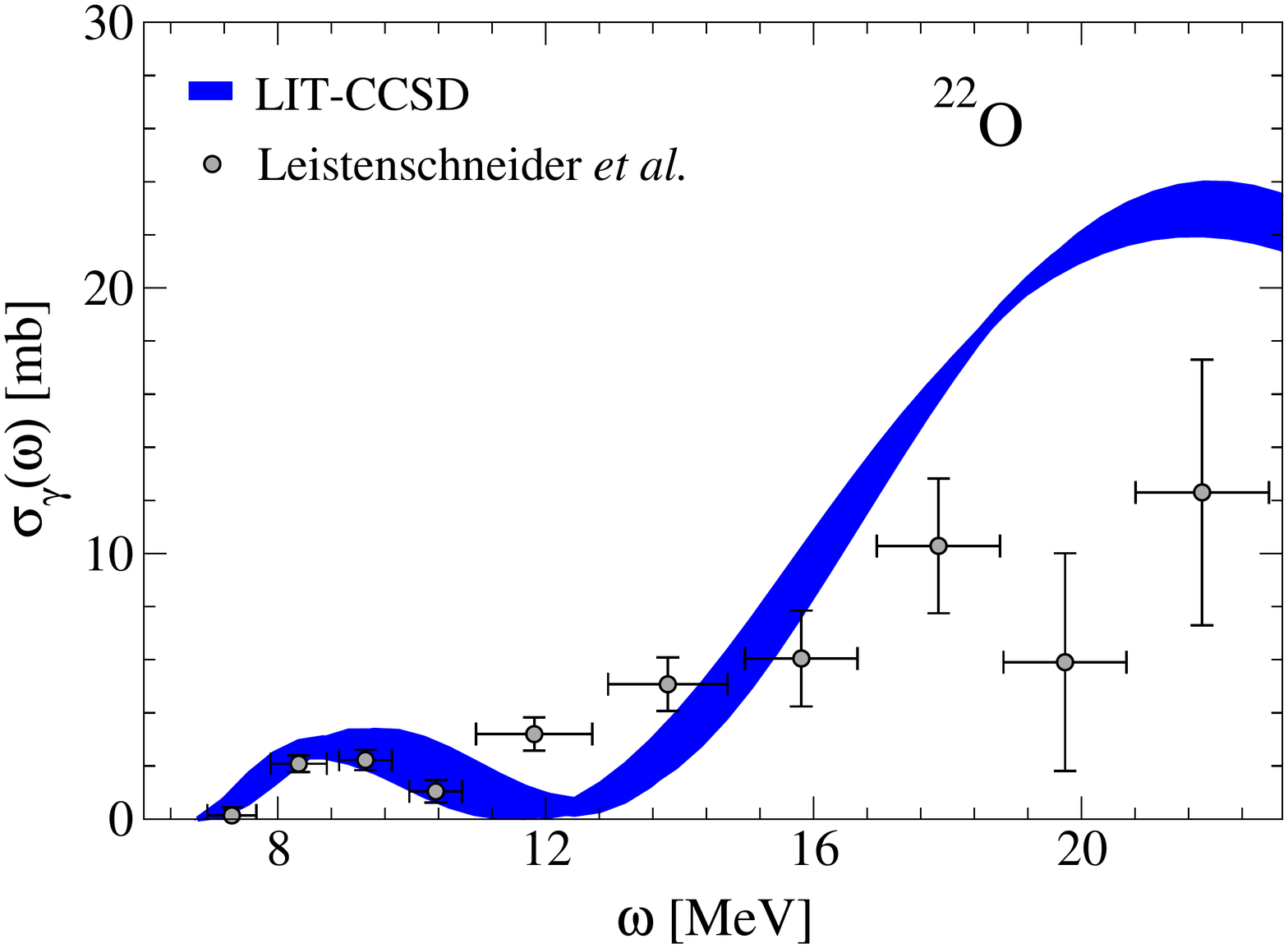}
  \caption{ $^{22}$O photodisintegration
  cross section compared with  data from Ref.~\cite{leistenschneider2001}.  The curve is calculated with the N$^3$LO nucleon-nucleon force and is shifted to the experimental threshold.}
  \label{fig_22O}
\end{minipage}
\hspace{.02\linewidth}
\begin{minipage}{.46\linewidth}
  \includegraphics[width=\linewidth]{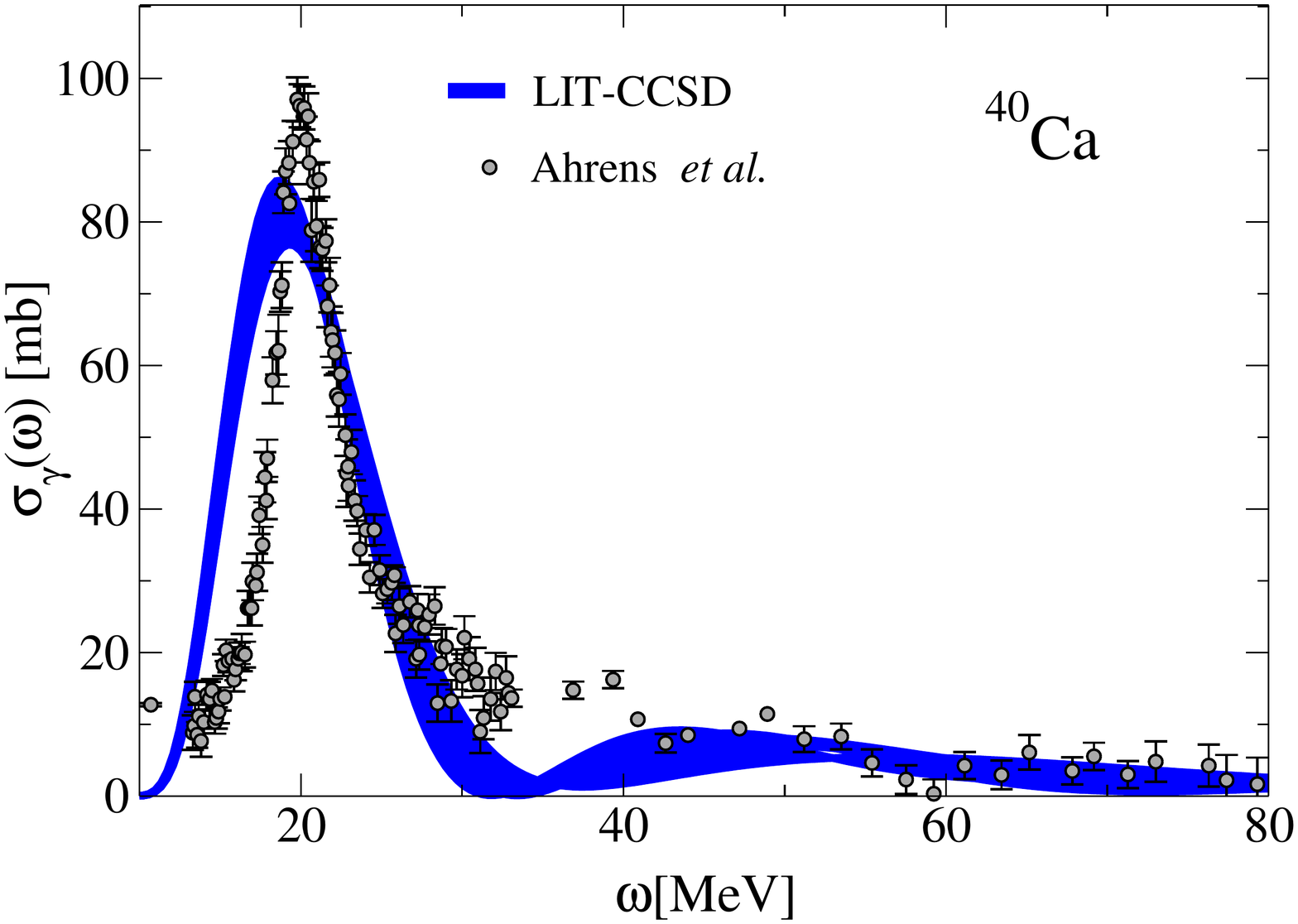}
  \caption{$^{40}$Ca photodisintegration
  cross section compared to  data from Ref.~\cite{Ahrens75}.  The curve is calculated with the N$^3$LO nucleon-nucleon force and is shifted to the  experimental threshold.}
  \label{fig_40Ca}
\end{minipage}
\end{figure}

In Refs.~\cite{PRL2013,PRC2014} we implemented the one-body operator of Eq.~(\ref{dipole}) in the solution of Eq.~(\ref{LITCC_eq}) and investigated the photodisintegration of a variety of light and medium-mass nuclei. 
After having benchmarked the LIT-CCSD method with exact hyperspherical harmonics~\cite{Barnea00} on $^4$He, where we have observed that higher-order corrections to the $T$ and ${\mathcal R}$ operator are very small, we have proceeded to study heavier nuclei. In particularly, here we review our results for the neutron-rich $^{22}$O isotope in Fig.~\ref{fig_22O} and the stable $^{40}$Ca nucleus in Fig.~\ref{fig_40Ca}.

For $^{22}$O one notices the
appearance of a small peak at low energy, also named pigmy dipole resonance and experimentally observed in a variety of neutron-rich nuclei, which agrees very nicely with data from Leistenschneider {\it et al.}~\cite{leistenschneider2001}. It is worth noticing that nothing in the interaction has been adjusted to $^{22}$O  and that no cluster structure has been imposed a priori in the calculation.
This is not the first time that the LIT approach suggests the
existence of a low-energy dipole mode, as in Ref.~\cite{bacca2002} a similar situation was found in 
 $^6$He. 

Because the computational cost of the coupled-cluster method scales mildly with respect to
the mass number, we can extend our studies to the photonuclear cross-section of  $^{40}$Ca. Fig.~\ref{fig_40Ca} shows a comparison of the LIT-CCSD calculations with data by Ahrens {\it et
  al.}~\cite{Ahrens75}. We clearly see  a giant resonance, even though the theoretical prediction is slightly broader and lower in strength. Both in  Fig.~\ref{fig_22O} and Fig.~\ref{fig_40Ca}
the band in the curve is obtained by inverting LIT with different $\Gamma$ parameters in Eq.~(\ref{lorenzo}) and can be seen as a lower estimate of the theoretical error bar.

\section{Electron-scattering reactions: Coulomb sum rule}
\label{sec-th}

\begin{figure}
\centering
\begin{minipage}{.43\linewidth}
  \includegraphics[width=\linewidth]{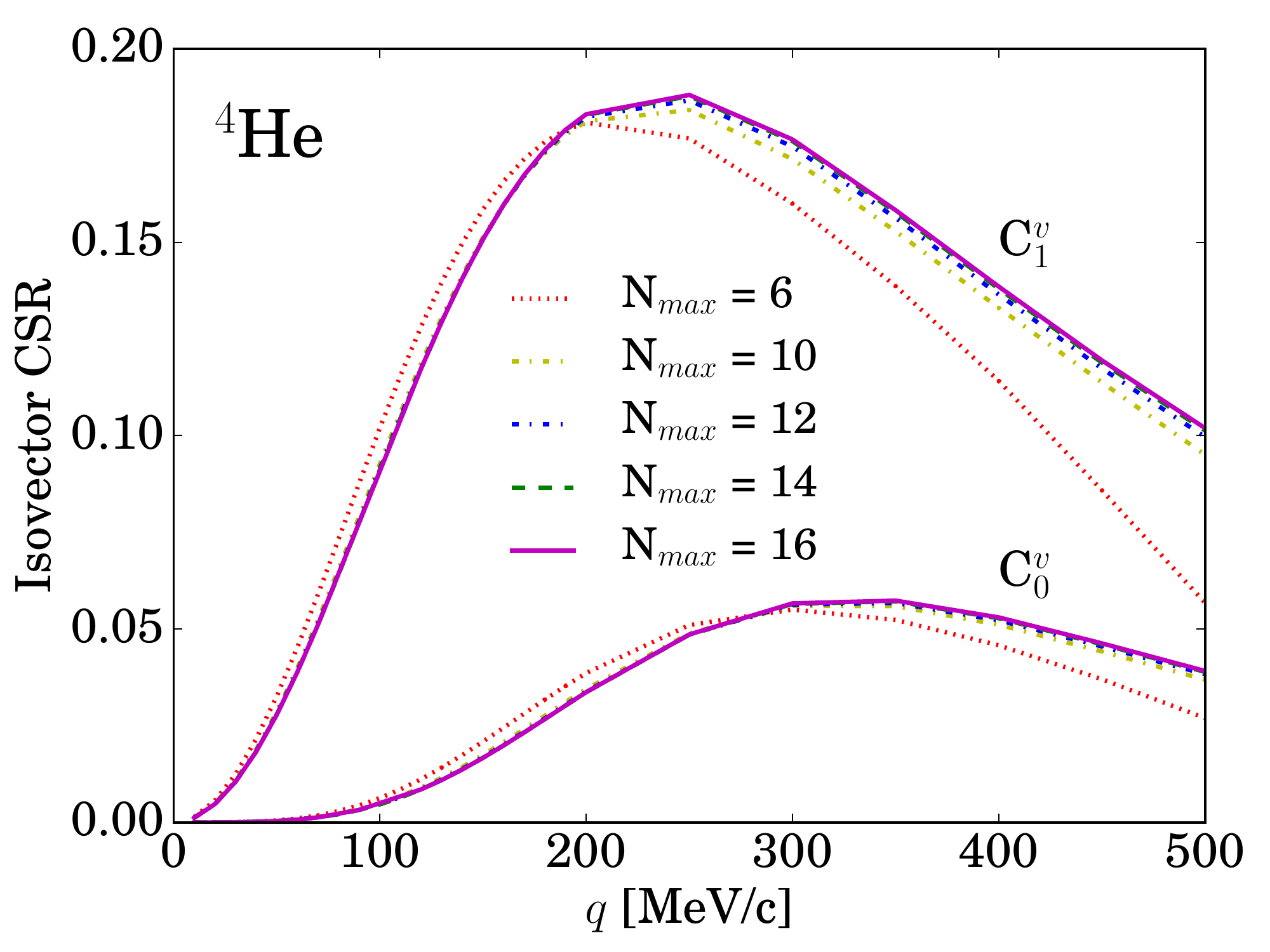}
  \caption{LIT-CCSD sum rules from $C^v_0$ and $C^v_1$ for $^4$He for different model space sizes $N_{max}$ with the N$^3$LO nucleon-nucleon chiral interaction.}
  \label{fig_convNmax}
\end{minipage}
\hspace{.02\linewidth}
\begin{minipage}{.43\linewidth}
  \includegraphics[width=\linewidth]{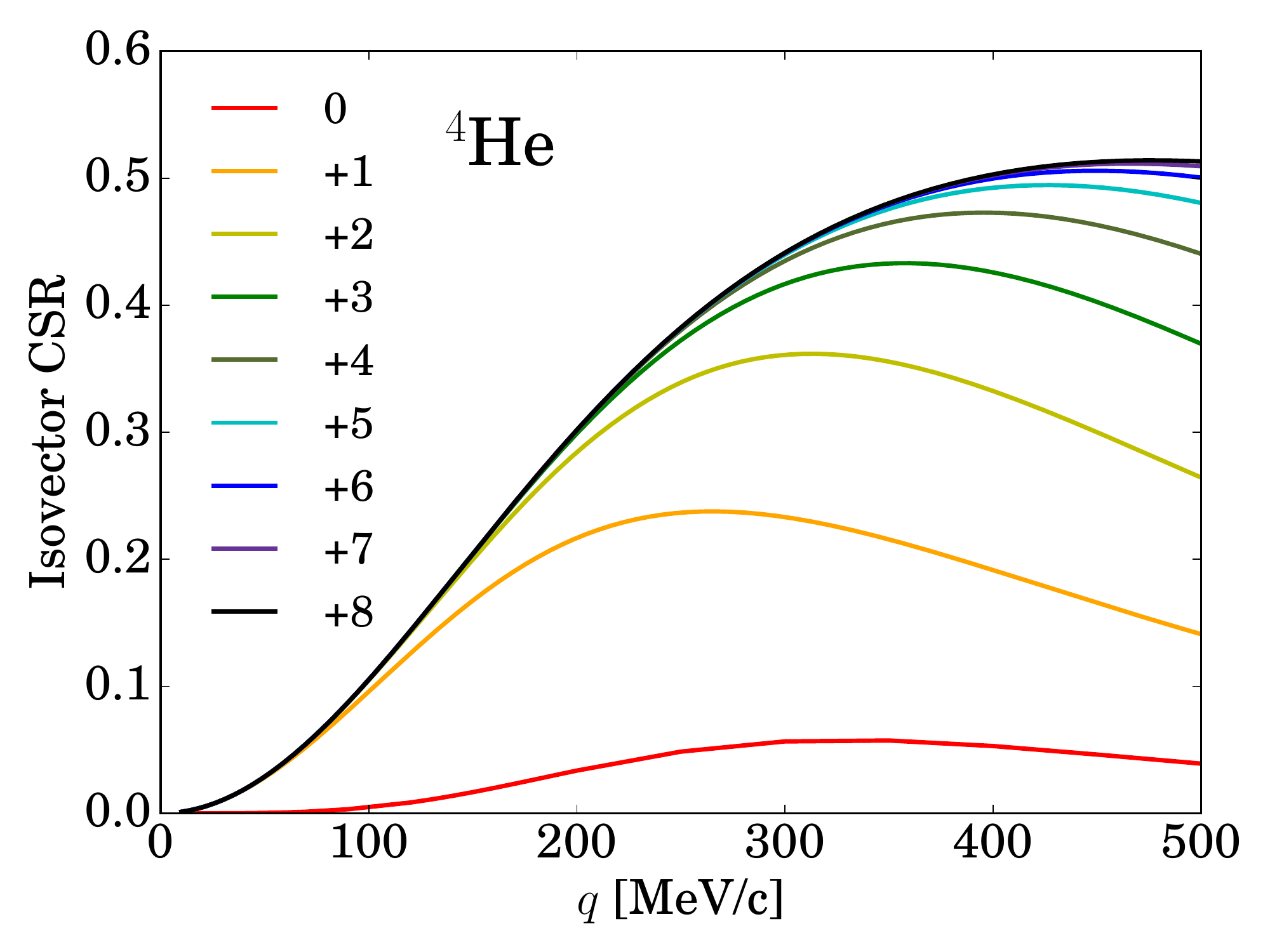}
  \caption{Recursive sum of isovector CSR for $^4$He calculated using the chiral N$^3$LO nucleon-nucleon potential in the LIT-CCSD scheme.}
  \label{fig_convJ}
\end{minipage}
\end{figure}

Electron scattering reactions have been already investigated within the LIT method, see, {\it e.g.}, \cite{th4he}.
With the objective in mind to eventually tackle neutrino-nucleus interactions relevant for long-baseline neutrino experiments,
we initiated a study of electron-scattering reactions within coupled-cluster theory. The latter has in fact the potential to be applied to $^{16}$O and other medium-mass nuclei, which are used as detector materials in neutrino experiments.
 Following  the work done with the Green's function Monte Carlo method in Ref.~\cite{Lovato}, we will first investigate the Coulomb sum rule.

The longitudinal Coulomb sum rule (CSR) is defined as
\begin{equation} 
S_L(q)=\frac{1}{Z}\int_{\omega^+_{th}}^\infty d\omega\frac{R_L(\omega,q)}{{G_E^{p}}^2(Q^2)},
\label{eq:lcsr}
\end{equation}
where $R_L(\omega,q)$ is the longitudinal response function, {\it i.e.},  Eq.~(\ref{eq:rs}) with
${\mathcal O}(q)=\rho(q)$ and a recoil term 
$\frac{\mathbf{q}^2}{2M}$ in the delta function, while $G_E^{p}(Q^2)$ is the proton electric form factor as a function
of the square of the four-momentum $Q^2$.
The charge operator can be written as
\begin{equation} 
\rho(q)=\sum_k^A\left(\frac{1+\tau_k^3}{2}G_E^p(Q^2)+\frac{1-\tau_k^3}{2}G_E^n(Q^2)\right)e^{i q z_k}\,,
\label{eq:CouOp}
\end{equation}
with $G_E^{n}(Q^2)$ being the neutron electric form factor.
Similarly to what done in Ref.~\cite{th4he} we perform a multipole expansion of Eq.~(\ref{eq:CouOp}) in terms of Coulomb multipoles, separating  isoscalar $(s)$  and isovector $(v)$ contributions as 
\begin{equation} 
\rho(q)=\sum_J \left (C_J^{s}(q)+C_J^{v}(q) \right) \,.
\label{eq:mult}
\end{equation}
Each of the Coulomb multipole is calculated separately  solving Eq.~(\ref{LITCC_eq}) for various  fixed values of  the momentum transfer $q$. Below, we show LIT-CCSD results for $^{4}$He obtained
using the  N$^3$LO nucleon-nucleon chiral interaction.

In Fig.~\ref{fig_convNmax} we test the  convergence of our calculations in terms of the model space size $N_{max}=2n+l$, determined by the number of single-particle shells used in the $ph$ expansion. For the isovector monopole $C^v_0$ and isovector dipole $C^v_1$ operator, one observes that very good convergence is reached already for $N_{max}=14$. For higher order multipoles, slightly larger model spaces are needed.
With respect to exact hyperspherical harmonics~\cite{Barnea00},  $C^v_1$  calculations from coupled-cluster theory agree very nicely up to about $q=200$~MeV/c, but are 20$\%$ smaller at higher momenta. Note that at energies of about 300 MeV and higher, the chiral interaction of Ref.~\cite{Entem03}  is at the limit of its applicability.
 Work is in progress to understand the role of center of mass contamination in our calculation, given that the operator in Eq.~(\ref{eq:CouOp}) is not translational invariant. 

In Fig.~\ref{fig_convJ}
we compare the recursive sum of the isovector multipoles that make up the CSR. One can readily see that, while at momentum transfer below $q=100$ MeV/c two or three multipoles suffice, at the highest momentum transfer considered $q=500$ MeV/c, nine multipoles need to be calculated before reaching convergence. A similar situation is found for the isoscalar multipoles.
\begin{figure}
\centering
\includegraphics[width=6cm,clip]{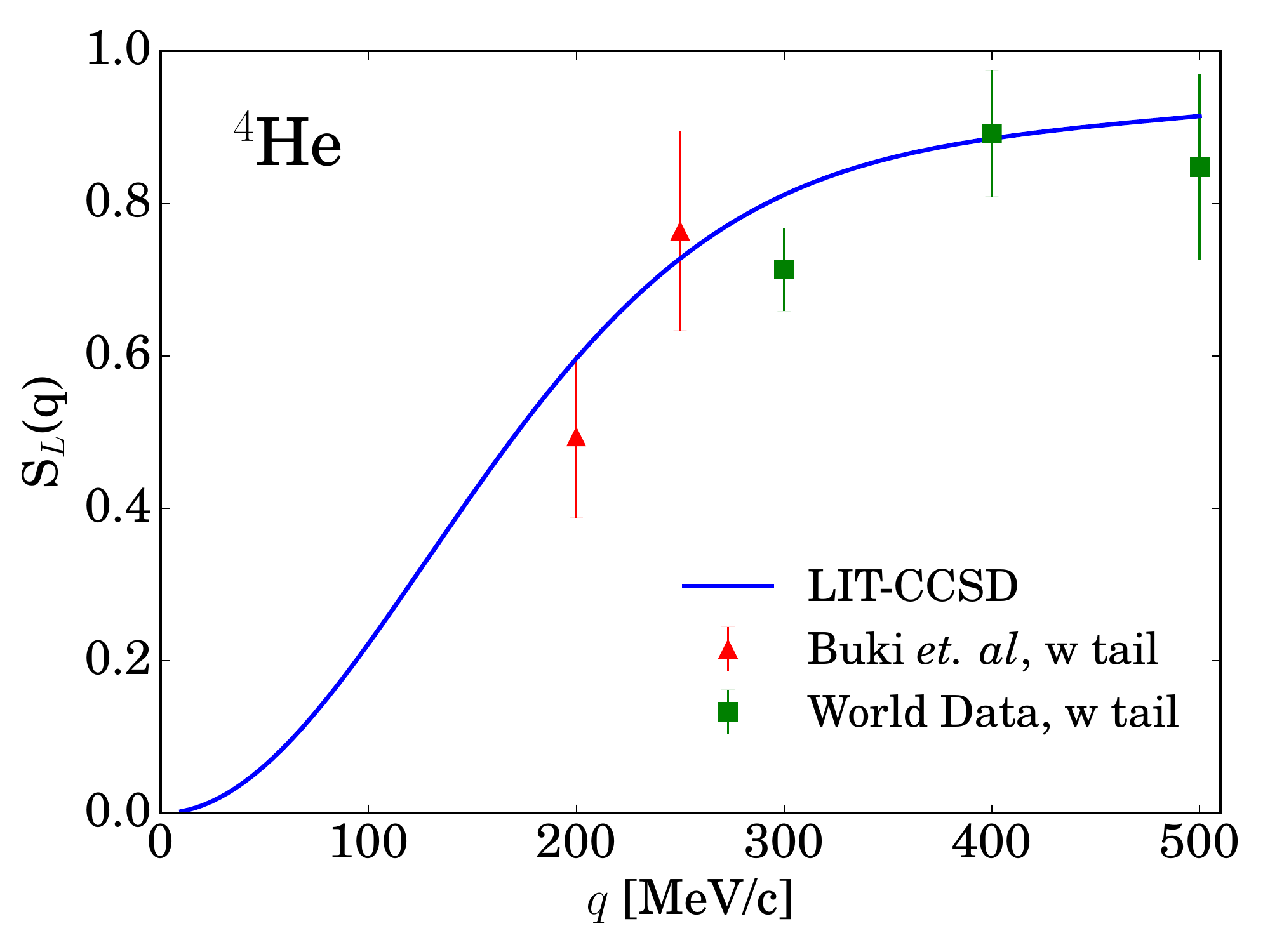}
\caption{Preliminary results for the $^4$He CSR from coupled-cluster theory compared with experimental CSR obtained from data from Buki {\it et. al.} (red triangle)~\cite{buki} and from Ref.~\cite{world_data} (green square), as described in the text.}
\label{fig-exp}       
\end{figure}
Finally, in Fig.~\ref{fig-exp} we compare the total CSR calculated from coupled-cluster theory with 
 experimental data.  Our calculation still misses the contribution of the isoscalar monopole term. Measurements of the longitudinal response functions at intermediate momentum transfer have been performed in the past and are collected in Ref.~\cite{world_data}, while low momentum data at $q=200$ and $250$ MeV/c are taken from Buki {\it et. al.}~\cite{buki}. Since finite maximal values of the energy transfer $\omega_{max}$ are measured in experimental data, the experimental CSR is obtained as
\begin{equation}
S_L(q)=\frac{1}{Z}\int_{\omega^+_{th}}^{\omega_{max}} d\omega\frac{R_L(\omega,q)}{G_E^{P^2}(Q^2)}+S_{L,\text{ tail}}
\end{equation}
where $S_{L,\text{ tail}}$ is taken from the theoretical calculations of $^4$He response functions of Ref.~\cite{th4he}.\\
As shown in Fig.~\ref{fig-exp}, our calculation agrees well with the experimental data from low- to intermediate-momenta. It is known from Ref.~\cite{th4he} that three-body forces are not very pronounced on the CSR at the values of momentum transfer where  data are available.  Work is being presently directed to benchmark these studies with hypersherical harmonics, with the prospect of tackling $^{16}$O.  

\section{Conclusions}
\label{sec-concl}
After a short review of the coupled-cluster formulation of LIT method, we report on our recent calculations of the photodisintegration of the neutron-rich $^{22}$O nucleus and the stable $^{40}$Ca medium-mass nucleus. We then show our preliminary calculation of the Coulomb sum rule for $^{4}$He, which describe rather well the available experimental data.
A benchmark with data is essential for any theory that is going to be used to model the neutrino-nucleus interaction taking place within the detectors of neutrino long-baseline experiments. Using the LIT-CCSD theory, our long term goal is to study neutrino
scattering off $^{16}$O, which is relevant for the T2K experiment.

\begin{acknowledgement}
We thank  N.~Barnea and T.~Papenbrock for useful discussions. This work was supported in parts by the Natural Sciences and Engineering Research Council (NSERC), the National Research Council of Canada, the Canadian Institute of Nuclear Physics (CINP), the U.S. Department of Energy (Oak Ridge National Laboratory), under Grant Nos. DEFG02-96ER40963 (University of Tennessee), de-sc0008499 (NUCLEI Sci-DAC collaboration), and the Field Work Proposal ERKBP57.
\end{acknowledgement}

%
%
%

\end{document}